\documentclass[seceq,supplement,preprint]{ptptex}

\usepackage{graphicx}
\preprintnumber{YITP-12-27}
\title{%        %You can use \\ for explicit line-break.
Quarkonium at $T>0$
}

\author{%       %Use \scshape for the family name.
Kenji \textsc{Morita}%
}

\inst{%     %Affiliation, neglected when [addenda] or [errata].
Yukawa Institute for Theoretical Physics, Kyoto University, Kyoto 606-8502 Japan
}

\abst{%         %This abstract is neglected when [addenda] or [errata].
We report recent progress on theoretical investigations of quarkonia at
finite temperature. We discuss medium modification of charmonia and
bottomonia from a viewpoint of local operators and
point out that while charmonia are sensitive to
the deconfinement transition bottomonia will be modified at much higher
temperatures.
}

\usepackage{amsmath}	% required for `\align' (yatex added)
\begin{document}

\maketitle

\section{Introduction}
Properties of heavy quarkonia have been extensively studied since
it was pointed out that they provide information on deconfinement
transition of QCD.\cite{Matsui-Satz,Hashimoto}  While expected suppressions of
the quarkonia have been measured in heavy ion experiments
\cite{experiment}, interpretation of these data is not so
straightforward, because of not only the complexity of the collision processes but also
the fact that spectral properties of the heavy quarkonia are not well
understood yet. In this report, we focus on recent development of
theoretical understanding of the quarkonium states at finite
temperature.

\section{Theoretical development}

Lattice QCD provides a unique non-perturbative first principle
approach. A direct study of the quarkonium spectral
function $\rho(\omega,T)$ is possible with the help of the maximum
entropy method (MEM) which
enables us to invert a current correlation function at imaginary time
$G(\tau,T)$ via a dispersion relation. \cite{Asakawa-Hatsuda}
\begin{equation}
 G(\tau,T)=\int d\omega
  \frac{\cosh[\omega(\tau-1/(2T))]}{\sinh[\omega/(2T)]} \rho(\omega,T).\label{eq:disp_lattice}
\end{equation}
 Even in recent calculations with bigger lattice sizes
 \cite{DingandNonaka}, however, quantitative information seems hard to be
 extracted, presumably due to limited temporal lattice size at high
 temperature and a small number of data points. The existence of the
 spectral peak does not necessary mean survival of a quarkonium\cite{MP} and substantial spectral modification does not
 contradict with the behavior of $G(\tau,T)$ \cite{morita_borel}.
A rather promising direction seems to determine an interquark
 potential $V(r,T)$ containing both real and imaginary parts for a
 Sch\"{o}dinger equation by lattice QCD.\cite{Alex}
 Indeed, existence of the imaginary part in the potential was pointed
 out by Laine et al. within a resummed perturbation theory.\cite{Laine}
 Recently significant progress has been made with an
 effective field theory framework for heavy quark bound
 states.\cite{pNRQCD} 
 Though this approach assumes a hierarchy of the energy scale which
 becomes complicated at finite temperature due to newly introduced
 medium energy scales, analytically
 results have been presented in some cases to give an insight of
 possible mechanisms of the in-medium modification of a
 quarkonium.\cite{weakcoupling}
 These analyses indicate that a dominant in-medium effect on quarkonium at
 experimentally accessible temperatures could be a singlet to octet
 breakup process by gluons. 
 On the other hand, an estimation of the energy density achieved in
 heavy ion collisions at RHIC energies and
 subsequent hydrodynamic evolution lead to a lifetime of the
 deconfined phase long enough to melt quarkonia if $\Gamma > 50$
 MeV.\cite{PHENIX} 
 This means that we need a theoretical estimation of the width at those temperatures, likely
 in the strongly interacting regime.\cite{HTL} 
 One of the promising approach for this purpose is relating local
 operators to a medium modification of a quarkonium via an operator product
 expansion (OPE). We will give a basic concept and a recent result
 below.

\section{Local operator approach for quarkonia}

\begin{figure}[ht]
 \includegraphics[width=0.45\textwidth]{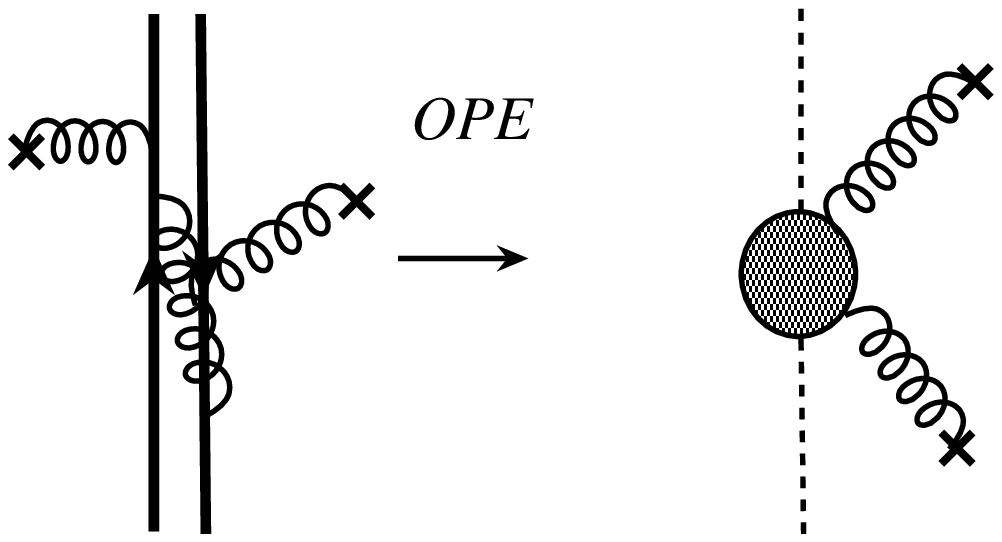}
 \includegraphics[width=0.45\textwidth]{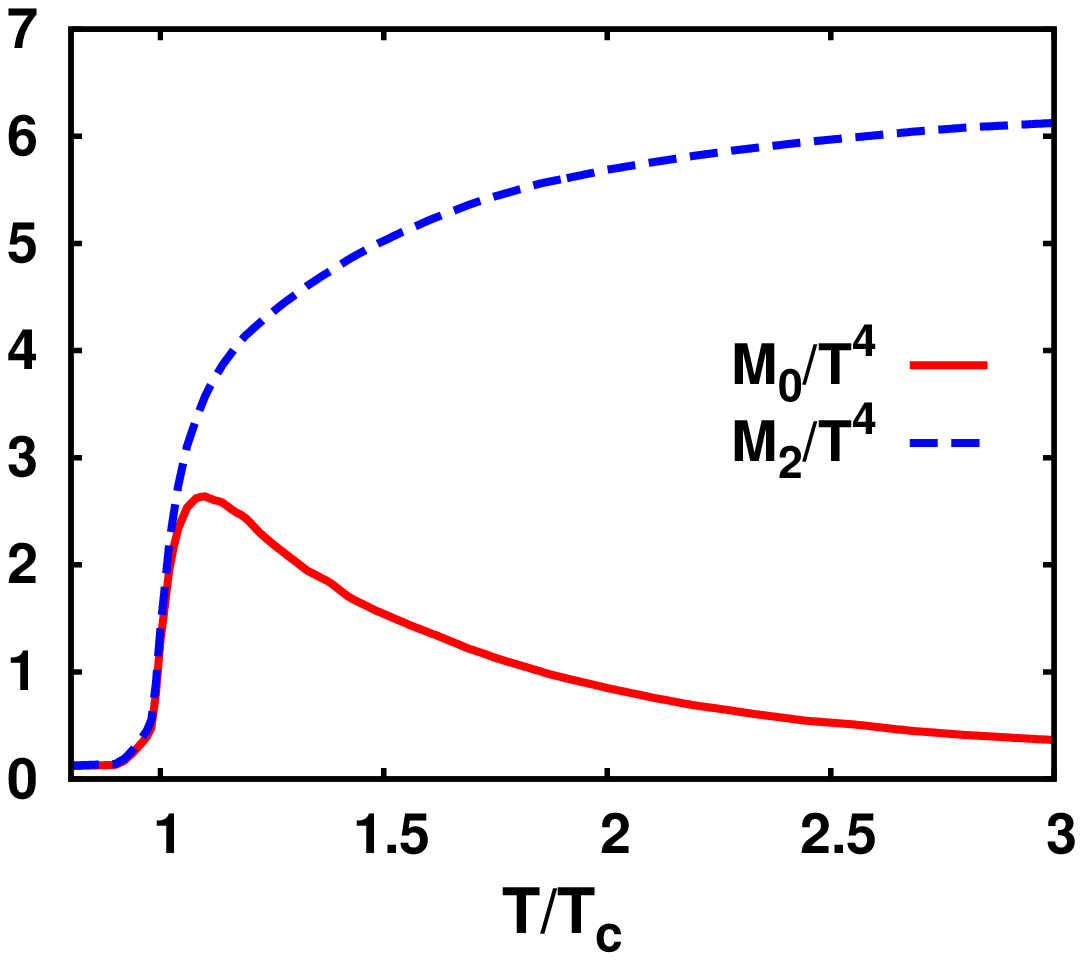}
 \caption{Left: Schematic diagram for OPE. Right: Temperature dependence of $M_0/T^4$ and $M_2/T^4$
 extracted from pure SU(3) lattice gauge theory. \cite{lattice_su3}}
 \label{fig:m0m2}
\end{figure}

The interaction between a heavy quarkonium and soft
gluons was first formulated by Peskin \cite{Peskin}.
The key concept is a separation scale, which is binding energy
$\epsilon$ in the
case of a heavy quarkonium. Regarding an exchange momentum $k$ larger
than $\varepsilon$ as hard scale while the other case as soft one, we may express a matrix
element via OPE as 
$\sum_{i} C_i \langle \mathcal{O}_i \rangle$ where $C_i$ and 
$\langle \mathcal{O}_i \rangle$
stand for the Wilson coefficients responsible for the hard scale 
$k > \epsilon$ and expectation value of local operators for the soft
scale, respectively (see Fig.~\ref{fig:m0m2}). 
The leading order
contribution is given by dimension four gluon condensate. At
lower temperature than the seperation scale, one may impose all the
medium effect on the change of the expectation value of the operators,
which can be extracted from lattice calculations. 
Taking the real part of the matrix element immediately leads to a formula of
a mass shift, which is the second order Stark effect in QCD, as
$ \Delta m_{\bar{Q}Q} =  -\frac{7\pi^2}{18}\frac{a_0^2}{\epsilon}\left\langle \frac{\alpha_s}{\pi}\Delta \boldsymbol{E}^2 \right\rangle$
where $a_0$ is the Bohr radius of the Coulombic bound state and 
$\left\langle \frac{\alpha_s}{\pi}\Delta \boldsymbol{E}^2 \right\rangle$
is the temperature dependent part of the chromoelectric condensate. 
The expectation value of the dimension four gluon operators can be
extracted from pressure $p$, energy density $\varepsilon$ and effective
coupling constant $\alpha_s^{\text{eff}}(T)$ obtained in lattice gauge theory
as
\begin{align}
 \left\langle \frac{\alpha_s}{\pi}\Delta\boldsymbol{E}^2 \right\rangle
 &=
 \frac{2}{11-\frac{2}{3}N_f}M_0(T)+\frac{3}{4}\frac{\alpha_s^{\text{eff}}(T)}{\pi}M_2(T)
\end{align}
where $M_0$ and $M_2$ correspond to the gluonic part of the QCD trace
anomaly, $\varepsilon-3p$ and that of enthalpy density, $\varepsilon+p$,
respectively and shown in Fig.~\ref{fig:m0m2}. The rapid change of the
energy density results in an abrupt increase of $\left\langle
\frac{\alpha_s}{\pi}\boldsymbol{E}^2\right\rangle$ thus downward mass
shift of a heavy quarkonium in the vicnity of the phase transition. 

The above framework with separation scale $\varepsilon$
is applicable for a deeply bounded heavy quark-antiquark
system. In reality, this condition might be questionable for charmonium
at high temperature. One can turn to the current correlation function
$\Pi(q^2)= \int d^4 x e^{iq\cdot x}\left\langle T j(x)j(0)\right\rangle$
in terms of OPE by going to deep Euclidean region in momentum space, 
$q^2 =-Q^2$.
Large negative $q^2$ enables us to compute $\Pi(q^2)$ in a perturbative manner with
a pointlike current such as $j^\mu(x) = \bar{c}(x)\gamma^\mu c(x)$ and makes
convergence property better than the above case. The relation to the
physical quarkonia, $q^2 = m_{\bar{Q}{Q}}^2$, is kept through the
dispersion relation. 
After the Borel transformation, which optimizes the dispersion relation such that the integral over the energy
is dominated by the lowest resonance,  the dispersion relation for the
transformed correlator $M(\nu)$ reads 
\begin{equation}
 \mathcal{M}(\nu) = \int dx^2 e^{-\nu x^2}\rho(2m_Qx, T)\label{eq:sumrule}
\end{equation}
where $\nu=4m_Q^2/M^2$ and $M$ is the so-called Borel mass parameter. 
QCD sum rules \cite{SVZ} gives a systematic framework to extract
spectral properties from the current correlation function and dispersion
relation also at finite temperature by introducing medium dependent condensates unless
the typical energy scale of the medium exceeds the separation scale.\cite{HKL} 
With a Breit-Wigner type pole ansatz for the physical spectral
function, one can derive a constraint on the spectral change of a
quarkonium at finite temperature \cite{morita_JpsiPRL,morita_borel}. 
Recently, Gubler and Oka proposed to apply MEM to QCD sum rules.\cite{MEMSumrule}
In this method,  we do not have to assume a specfic
functional form on the spectral function. Futhermore, compared to
lattice calculations based on Eq.~\eqref{eq:disp_lattice}, we can take as
many points as possible and the dispersion relation does not have a
temperature dependence other than the spectral function itself. 
\begin{figure}[ht]
 \includegraphics[width=0.45\textwidth]{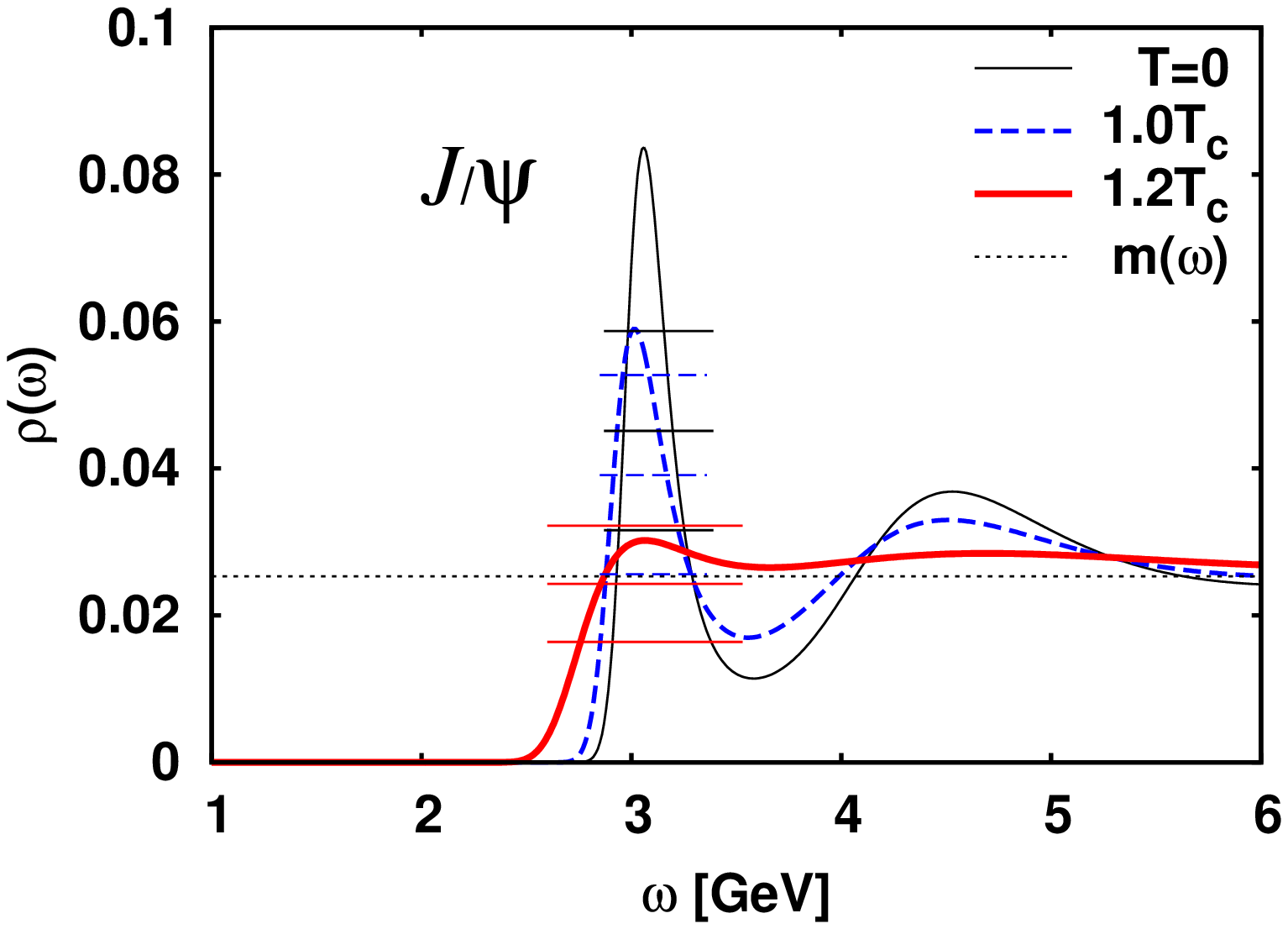}
 \includegraphics[width=0.45\textwidth]{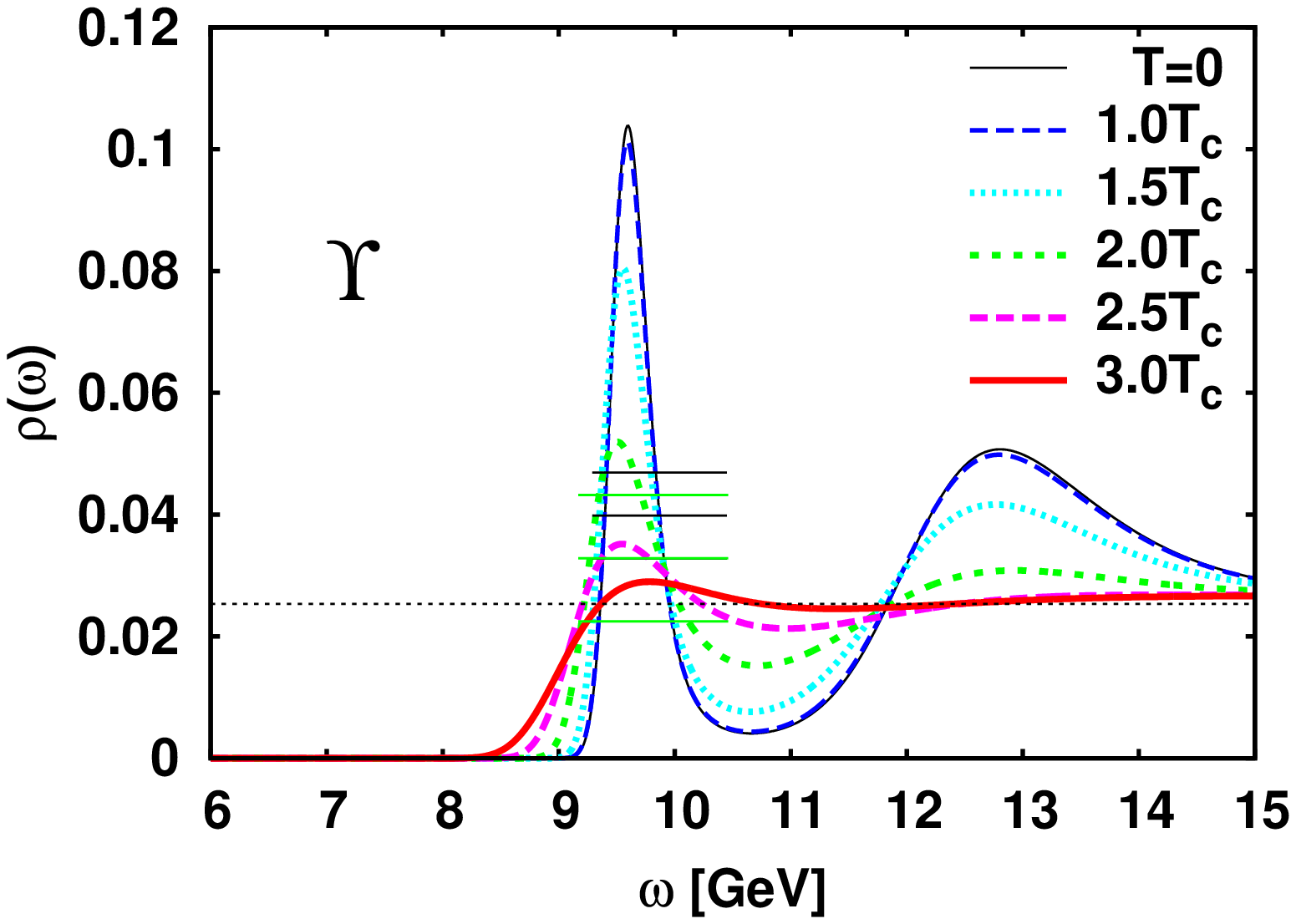}
 \caption{Spectral function obtained from QCD sum rules with maximum
 entropy method. Left: $J/\psi$, Right: $\Upsilon$. }
 \label{fig:mem}
\end{figure}
Figure \ref{fig:mem} shows spectral functions of
$J/\psi$ (left)\cite{Gubler_PRL} and
$\Upsilon$ (right)\cite{Bottom_MEM} obtained by the QCD sum rule+MEM approach.
 Although resolution of the width in the lowest peak is not sufficient,
one sees how the peak dissolves as temperature increases. The drastic
change around $T_c$ seen in the case of $J/\psi$ is consistent with previous sum rule
calculations \cite{morita_JpsiPRL,morita_borel}, while $\Upsilon$ hardly
reflects the phase transition but exhibits sizable modification above $2T_c$. 
\section{Summary and outlook}
Several approaches have revealed that a heavy quarkonium has a width at
finite temperature induced by the medium effect, especially due to
gluonic dissociations. Utilizing OPE, we have shown that charmonia are
sensitive to the phase transition while bottomonium are modified at much
higher temperatures. Since bottomonia will be more appropriate for
treatment with effective field theories, we expect theoretical attempts
could meet together to unveil the interaction of the heavy quarkonium in
the deconfined medium, which can be explored in heavy ion colllisions at the LHC energies.

\section*{Acknowledgements}
The author is much indebt to S.~H.~Lee for a fruitful collaboration.
He is also grateful to P.~Gubler, M.~Oka and K.~Suzuki for
results shown in Fig.~\ref{fig:mem}. This work is supported by YIPQS at
Kyoto University.

%\appendix
%\section{First Appendix} %Empty argument \section{} yields `Appendix'. 
%
%\section{Second Appendix}


\begin{thebibliography}{99}
%%%%%%%%%%%%%%%%%%%%%%%%%%%%%%%%%%%%%%%%%%%%%%%%%%%%%%%%%%%%%
% Some macros are available for the bibliography:
%  o for general use
%    \JL : general journals                 \andvol : Vol (Year) Page
%  o for individual journal 
%    \AJ   : Astrophys. J.           \NC         : Nuovo Cim.
%    \ANN  : Ann. of Phys.           \NPA, \NPB  : Nucl. Phys. [A,B]
%    \CMP  : Commun. Math. Phys.     \PLA, \PLB  : Phys. Lett. [A,B]
%    \IJMP : Int. J. Mod. Phys.      \PRA - \PRE : Phys. Rev. [A-E]     
%    \JHEP : J. High Energy Phys.    \PRL        : Phys. Rev. Lett.
%    \JMP  : J. Math. Phys.          \PRP        : Phys. Rep.
%    \JP   : J. of Phys.             \PTP        : Prog. Theor. Phys.     
%    \JPSJ : J. Phys. Soc. Jpn.      \PTPS       : Prog. Theor. Phys. Suppl.
% Usage:
%  \PRD{45,1990,345}          ==> Phys.~Rev.\ D \textbf{45} (1990), 345
%  \JL{Nature,418,2002,123}   ==> Nature \textbf{418} (2002), 123
%  \andvol{123,1995,1020}    ==> \textbf{123} (1995), 1020
%%%%%%%%%%%%%%%%%%%%%%%%%%%%%%%%%%%%%%%%%%%%%%%%%%%%%%%%%%%%%
  
\bibitem{Matsui-Satz}
	T.~Matsui and H.~Satz, \PLB{178,1986,416}
\bibitem{Hashimoto}
	T.~Hashimoto, O.~Miyamura, K.~Hirose and T.~Kanki, \PRL{57,1986,2123}
\bibitem{experiment}	
	R.~Granier de Cassagnac, these proceedings.
\bibitem{Asakawa-Hatsuda}
	M.~Asakawa and T.~Hatsuda, \PRL{92,2004,012001}
\bibitem{DingandNonaka}
	H.~-T.~Ding, A.~Francis, O.~Kaczmarek, F.~Karsch, H.~Satz and
	W.~Soeldner, arXiv:1107.0311;
	C.~Nonaka, M.~Asakawa, T.~Hoshino, M.~Kitazawa and Y.~Kohno, PoS
	\textbf{LAT2010} (2010) 207.
 \bibitem{MP}
	 \'A.~M\'{o}csy and P.~Petreczky, \PRD{77,2008,014501}.
 \bibitem{morita_borel}
	 K.~Morita  and S.~H.~Lee, \PRD{82,2010,054008}.
 \bibitem{Alex}
	 A.~Rothkopf, T.~Hatsuda, and S.~Sasaki, arXiv:1108.1579.
 \bibitem{Laine}
	 M.~Laine, O.~Philipsen, M.~Tassler and P.~Romatschke,
	 \JHEP{0703,2007,054}.
 \bibitem{pNRQCD}
	 N.~Brambilla, J.~Ghiglieri, A.~Vairo and P.~Petreczky,
	 \PRD{78,2008,014017}.
 \bibitem{weakcoupling}
	 N.~Brambilla, M.~A.~Escobedo, J.~Ghiglieri, J.~Soto and
	 A.~Vairo, \JHEP{1009,2010,038}.
 \bibitem{PHENIX}
	 See e.g., K.~Adcox et al., (PHENIX Collaboration),
	 \NPA{757,2005,184}.
 \bibitem{HTL}
	 J.~O.~Andersen, L.~E.~Leganger, M.~Strickland and N.~Su, \JHEP{1108,2011,053}.

 \bibitem{Peskin}
	 M.~E.~Peskin, \NPB{156,1979,365}.

 \bibitem{lattice_su3}
	 G.~Boyd et al., \NPB{469,1996,419}

 \bibitem{stark}
	 S.~H.~Lee and K.~Morita, \PRD{79,2009,011501}

 \bibitem{SVZ}
	 M.~A.~Shifman, A.~I.~Vainshtein and V.~I.~Zakharov,
	 \NPB{147,1979,385},448.

 \bibitem{HKL}
	 T.~Hatsuda, Y.~Koike and S.~H.~Lee, \NPB{394,1993,221}.

 \bibitem{morita_JpsiPRL}
	 K.~Morita and S.~H.~Lee, \PRL{100,2008,022301}.

 \bibitem{MEMSumrule}
	 P.~Gubler and M.~Oka, \PTP{124,2010,995}.

 \bibitem{Gubler_PRL}
	 P.~Gubler, K.~Morita and M.~Oka, \PRL{107,2011,092003}.

 \bibitem{Bottom_MEM}
	 P.~Gubler, K.~Suzuki, K.~Morita and M.~Oka, in preparation.
\end{thebibliography}
\end{document}